\documentclass[prb,aps,twocolumn,groupedaddress]{revtex4}

\usepackage{graphicx}  
\usepackage{dcolumn}   
\usepackage{bm}        
\usepackage{amssymb}   
\usepackage{amsmath}   
\usepackage{mathtools} 
\usepackage{natbib}    
\usepackage{lipsum}
\setcitestyle{super}

\begin{document}

\title{Changing the lattice position of a bistable single magnetic dopant in a semiconductor using a scanning tunneling microscope}

\author{J. M. Moore} \affiliation{Department of Physics \& Astronomy and Optical Science and Technology Center, University of Iowa, Iowa City, IA 52242}
\author{V. R. Kortan} \affiliation{Department of Physics \& Astronomy and Optical Science and Technology Center, University of Iowa, Iowa City, IA 52242}
\author{M. E. Flatt\'e} \affiliation{Department of Physics \& Astronomy and Optical Science and Technology Center, University of Iowa, Iowa City, IA 52242}
\author{J. Bocquel} \affiliation{Department of Applied Physics, Eindhoven University of Technology, Eindhoven, The Netherlands}
\author{P. M. Koenraad} \affiliation{Department of Applied Physics, Eindhoven University of Technology, Eindhoven, The Netherlands}

\begin{abstract}

We report a reversible and hysteretic change in the topography measured with a scanning tunneling microscope near a single Fe dopant in a GaAs surface when a small positive bias voltage is applied. 
First-principles calculations of the formation energy of a single Fe atom embedded in GaAs as a function of displacement from the substitutional site support the interpretation of a reversible lattice displacement of the Fe dopant. Our calculations indicate a second stable configuration for the Fe dopant within the lattice, characterized by a displacement along the [111] axis, accompanied by a change in atomic configuration symmetry about the Fe from four-fold to six-fold symmetry. The resulting atomic configurations are then used within a tight-binding calculation to determine the effect of a Fe position shift on the topography.  These results expand the range of demonstrated local configurational changes induced electronically for dopants, and thus may be of use for sensitive control of  dopant properties and dopant-dopant interactions.

\end{abstract}

\maketitle

\section{Introduction}

The manipulation of single substitutional impurities in a host semiconductor has become increasingly important in the development of nanoscale semiconductor devices, permitting a wide range of tunable phenomena. Recent advances in nanofabrication techniques allow the precise placement of an individual dopant atom in a host lattice~\cite{Schofield2003}, as well as the construction of semiconductor device sizes such that a single atomic impurity can significantly affect the electronic, magnetic, and optical properties of the host~\cite{Koenraad2011}, or form the active component of the device~\cite{Fuechsle2012}. Scanning tunneling microscopes (STM) allow remarkable precision and detail that is ideal for investigating the electronic structures of these solotronic devices, and have been successfully used to map their spatial structures and wave functions with atomic resolution~\cite{Yakunin2004b}. Recently, electronic control of the lattice position of individual dopants using a STM has been demonstrated,  from the reversible displacement of a single Si dopant in the surface layer of GaAs from a substitutional to an interstitial site~\cite{Garleff2011,Yi2011}, with a corresponding change of the dopant charge state. This transition is similar to the naturally occurring formation\cite{Chadi1989} of a $DX^-$ center in Al$_{x}$Ga$_{1-x}$As alloys with $x\geq0.22$.

\begin{figure}
\includegraphics[width=.8\columnwidth]{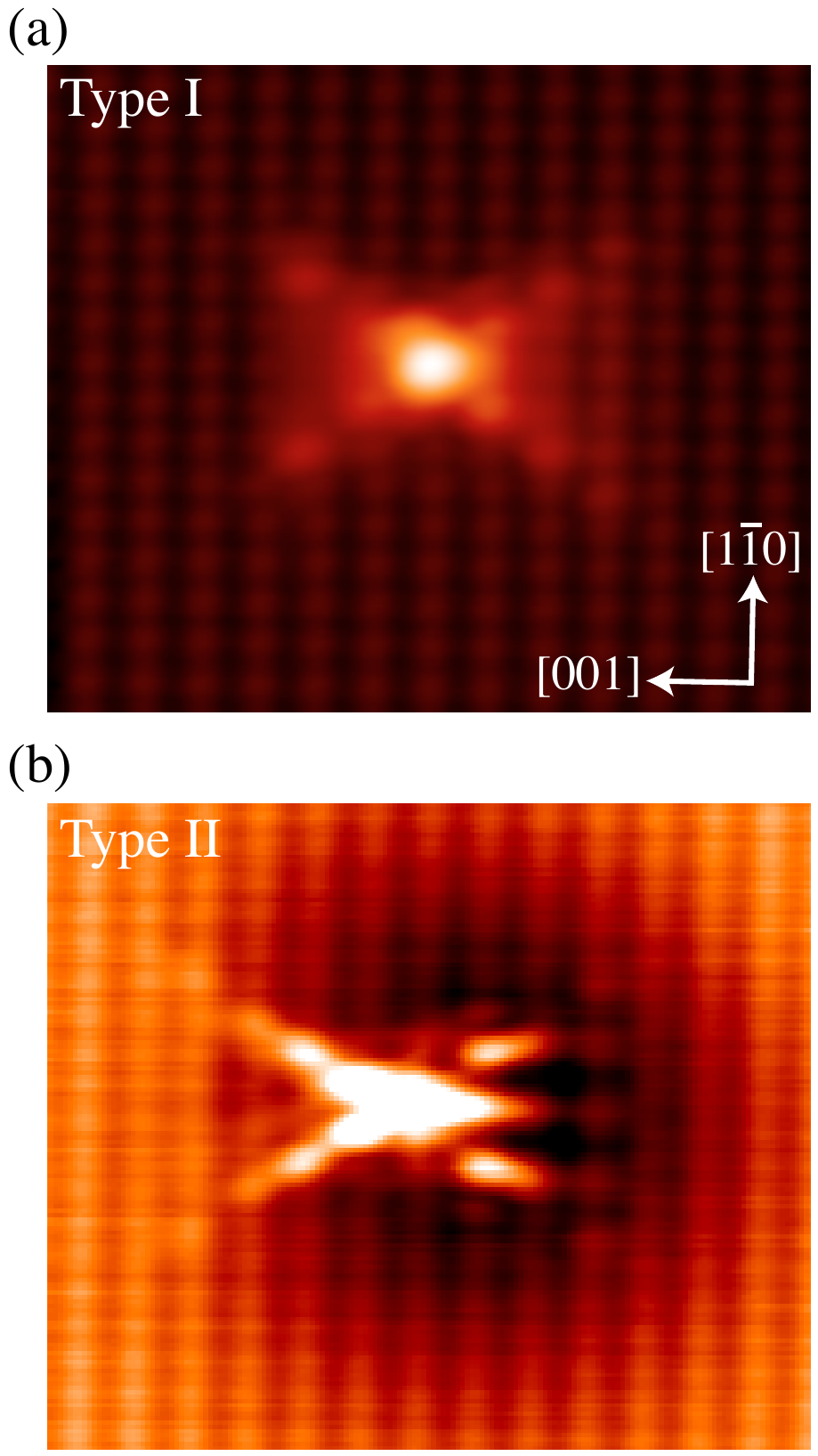}
\caption{\label{types} STM topography images of the same Fe atom that is located about 3 ML below the (110) surface of GaAs before and after switching. In (a) the expected (type~I) contrast for Fe in the isoelectronic 3$d^5$ (Fe$^{3+}$)$^0$ configuration is shown whereas in (b) type~II contrast after the switching is shown. Both type~I  and type~II topography were observed at the same small positive bias voltages. The type~I topography is perfectly symmetric with respect to the [001] axis and highly symmetric with respect to the [$1\bar10$] axis.  The type~II topography consists of a bright anisotropic feature on a dark background.  The dark background is attributed to the Coulomb field of negative charge located near the center of the Fe impurity in the type~II state. Image size 7.5 nm by 8.7 nm}
\end{figure}

Magnetic, transition metal (TM) dopants in III-V semiconductors are of considerable interest due to the internal spin degrees of freedom associated with the core $d$-electron states. Control of the core occupation of individual TM atoms would make it possible to manipulate the core spin and sensitive magnetic interactions of the dopants. In recent work\cite{Bocquel2013} the manipulation of the charge state of single Fe dopants at the [110] surface of GaAs using cross-sectional scanning tunneling microscopy (X-STM) was demonstrated. Direct electrical manipulation of the core occupation of a single Fe dopant from the isoelectronic 3$d^5$ (Fe$^{3+}$)$^0$ configuration to the ionized acceptor state 3$d^6$ (Fe$^{2+}$)$^-$ was achieved using tip-induced band bending (TIBB) by the STM, changing the core spin state from S=5/2 to S=2.  Under the conditions of Ref.~\onlinecite{Bocquel2013}, for a small positive bias voltage of the STM tip (where we normally observe a contrast  related to the 3$d^5$ (Fe$^{3+}$)$^0$ configuration, see Fig.~\ref{types}(a)) we frequently experienced  a sudden switch of the contrast. The contrast after switching is shown in Fig.~\ref{types}(b). The rather peculiar electronic contrast of the topography in the proximity of the dopant consists of a bright strongly anisotropic feature superimposed on a dark symmetric background. 

\begin{figure}
\includegraphics[width=\columnwidth]{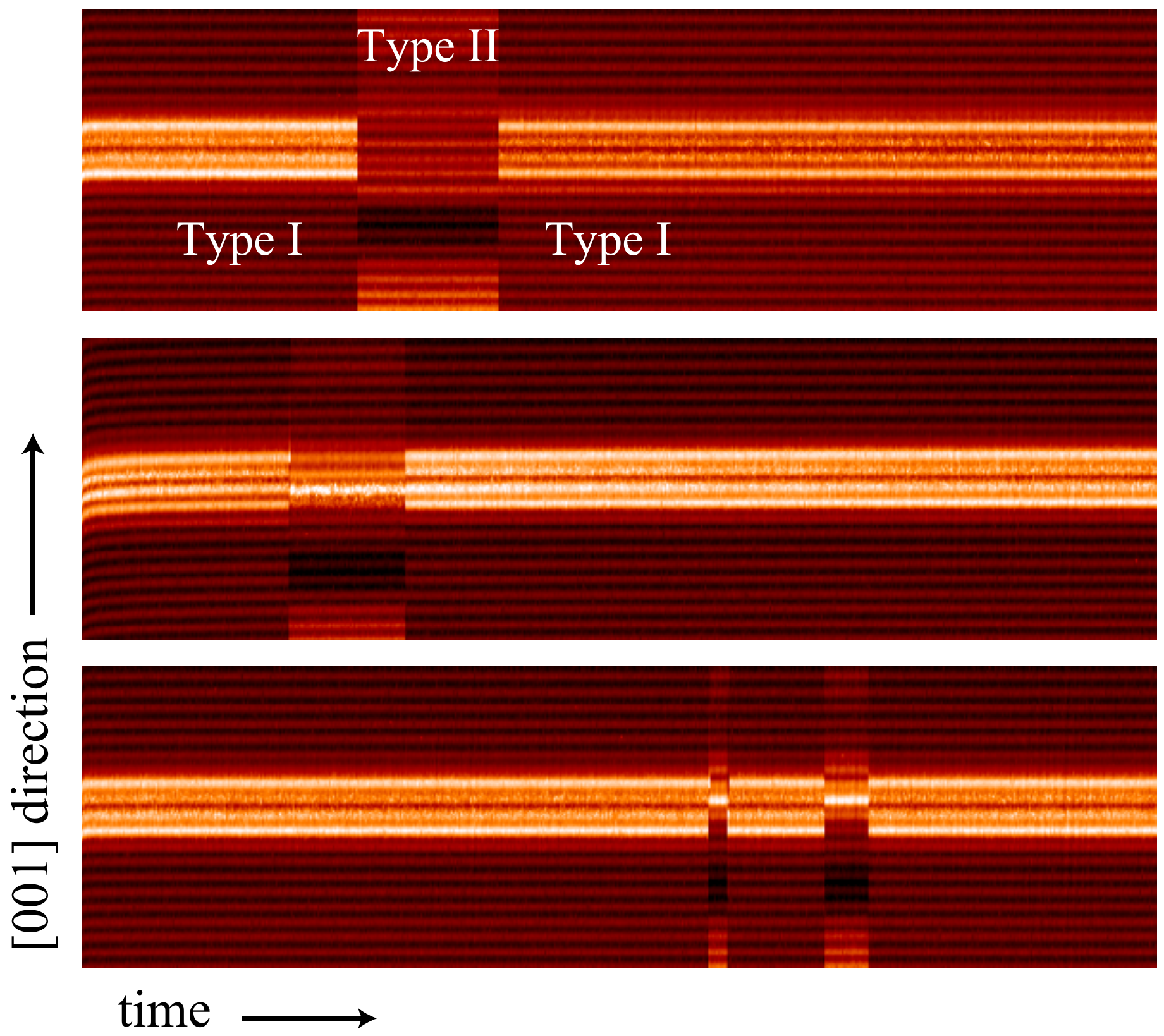}
\caption{\label{scan} Time dependence of the switching behavior of individual Fe atoms. The time traces are obtained by repeatedly scanning over the same line intersecting with the topographic contrast of a Fe atom. In each panel the tip was repeatedly scanning along the same line in the [001] direction but the line is located at a different position in the [1-10] direction with respect to the position of the Fe atom. The length of the time traces is 120 seconds and the length of the scan line in the [001] direction is 9.5 nm.}
\end{figure}

After a brief period, typically of $\sim 10$ seconds, the contrast switches back to the original contrast (which we refer to as type~I) shown in Fig.~\ref{types}(a). We followed the switching behavior of single Fe atoms by repeatedly scanning over the same scan line in the [001] that is intersecting with the contrast of the Fe impurity. In Fig.~\ref{scan} several time-traces are shown that consist of repeated scan lines taken in the [001] direction which are plotted next to each other which demonstrate the time dependence of the switching process. In all panels  the switching appears to occur instantaneously, and  the switch to the type~II contrast shown in Fig.~\ref{types}(b) lasts for about 10 seconds. The dark background observed for the type~II state we interpret as due to negative charge that is localized at the scale of single atom, which affects the tunneling current\cite{Garleff2011,Teichmann2008}.

The shift seen in topography is not associated with a shift in the charge state alone. In Ref.~\onlinecite{Bocquel2013} by  modifying the occupation of the core $d$-level  the charge state of the Fe atom was changed to 3$d^6$ (Fe$^{2+}$)$^-$. This negative charge state binds a hole in an effective-mass  state with a binding energy of 25~meV \cite{Malguth2008}. If the bright anisotropic contrast for Fe in the type~II state were due to a hole  bound by 25~meV then a much larger scale and triangular shaped contrast should be visible, such as that associated with the 25~meV bound state of Zn\cite{Loth2006,Celebi2010}, which is not the case. The bright anisotropic contrast suggests a much deeper binding energy and therefore  a relaxation of the lattice position of the Fe dopant from substitutional to interstitial lattice position may be responsible for this observation. Such a switching scenario is similar to that which is seen in the case of Si in the surface layer of GaAs, in which the configuration of the Si atom changes from substitutional to interstitial with an associated change of charge state~\cite{Garleff2011,Yi2011}.

The existence of a second possible configuration would require two minima in the potential energy landscape in configuration space separated by a potential barrier. For the case of Si in GaAs, the configuration associated with the second minimum is described by a change of the charge state from the neutral $sp^3$ (Si$^{2+}$)$^0$ state to the negatively charged $sp^2$ (Si$^+$)$^-$ state.  In addition, the charge density is strongly localized near the Si atom, causing a structural relaxation corresponding to an approximate 1 \AA\ displacement along the [111] axis (along the axis of the Si-As bond) and away from the 4-fold symmetric substitutional position, resulting in a broken bond~\cite{Chadi1989}. At the GaAs [110] surface, the formation energy of the interstitial configuration is negative and stable, while in bulk GaAs the formation energy is positive and metastable~\cite{Yi2011}. To evaluate the possibility of a reversible, electronically-induced shift in the lattice position of the Fe dopant we use first-principles calculations to investigate the electronic structures for both cases of substitutional Si and Fe dopants in bulk GaAs and their associated formation energy profiles in configuration space.

\section{Theoretical \& Computational Methods}

Following the previous work on the energetics of the $DX^-$ center by Chadi and Chang~\cite{Chadi1989}, the $DX^-$ center is formed when an impurity in the 4-fold symmetric substitutional site, denoted by $d$, which gives up one electron to the conduction band,

\begin{figure*}
\includegraphics[width=\textwidth]{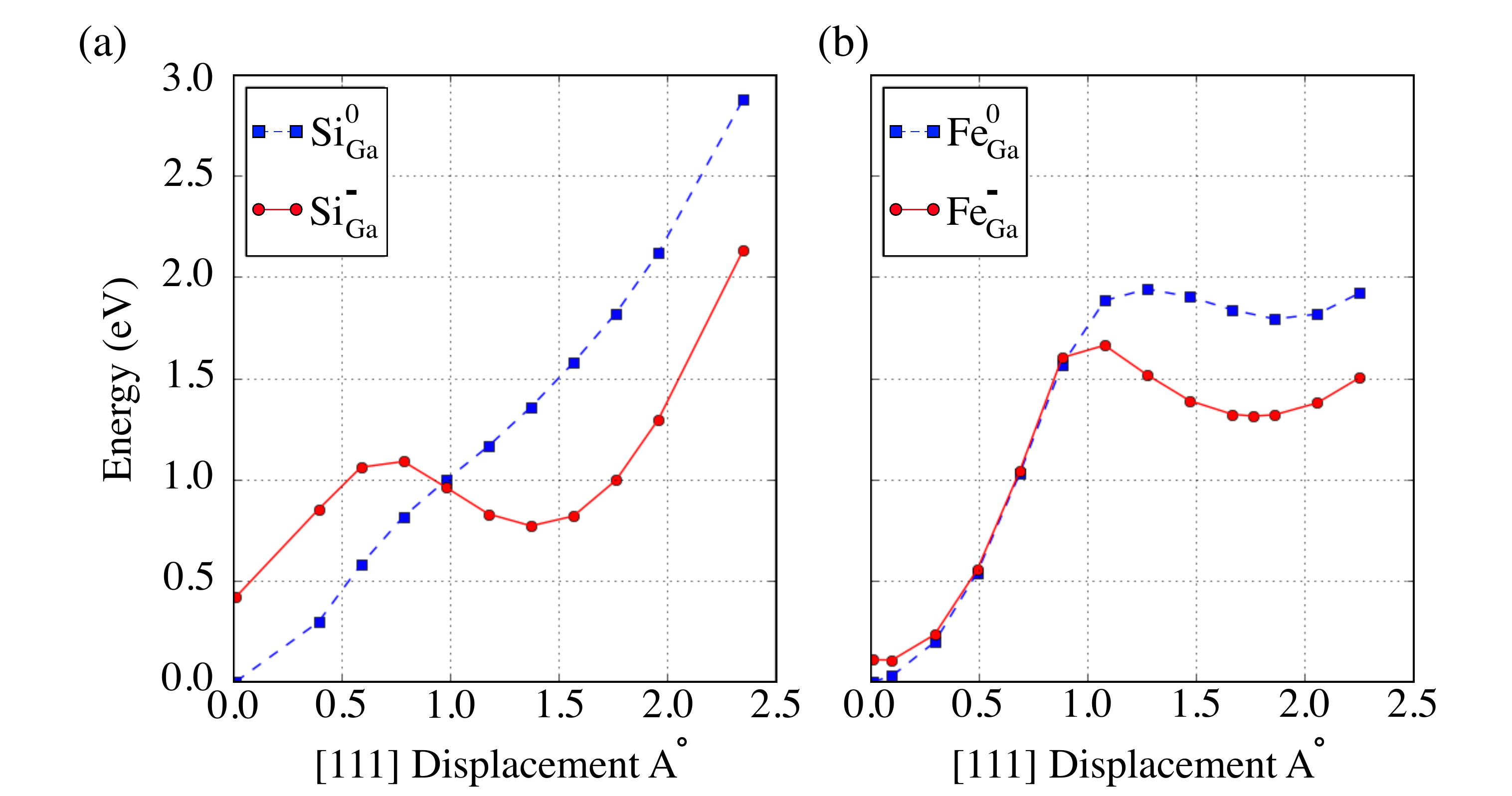}
\caption{\label{energies}Energies of Si (a) and Fe (b) in configuration space embedded in bulk GaAs for the neutral and negative charge states. For convenience, the zero of energy is taken to be the total energy of the neutral charge state with the dopant atom in the 4-fold symmetric substitutional configuration, and the displacements along the [111] axis are relative to the substitutional configuration.}
\end{figure*}

\begin{figure}
\includegraphics[width=\columnwidth]{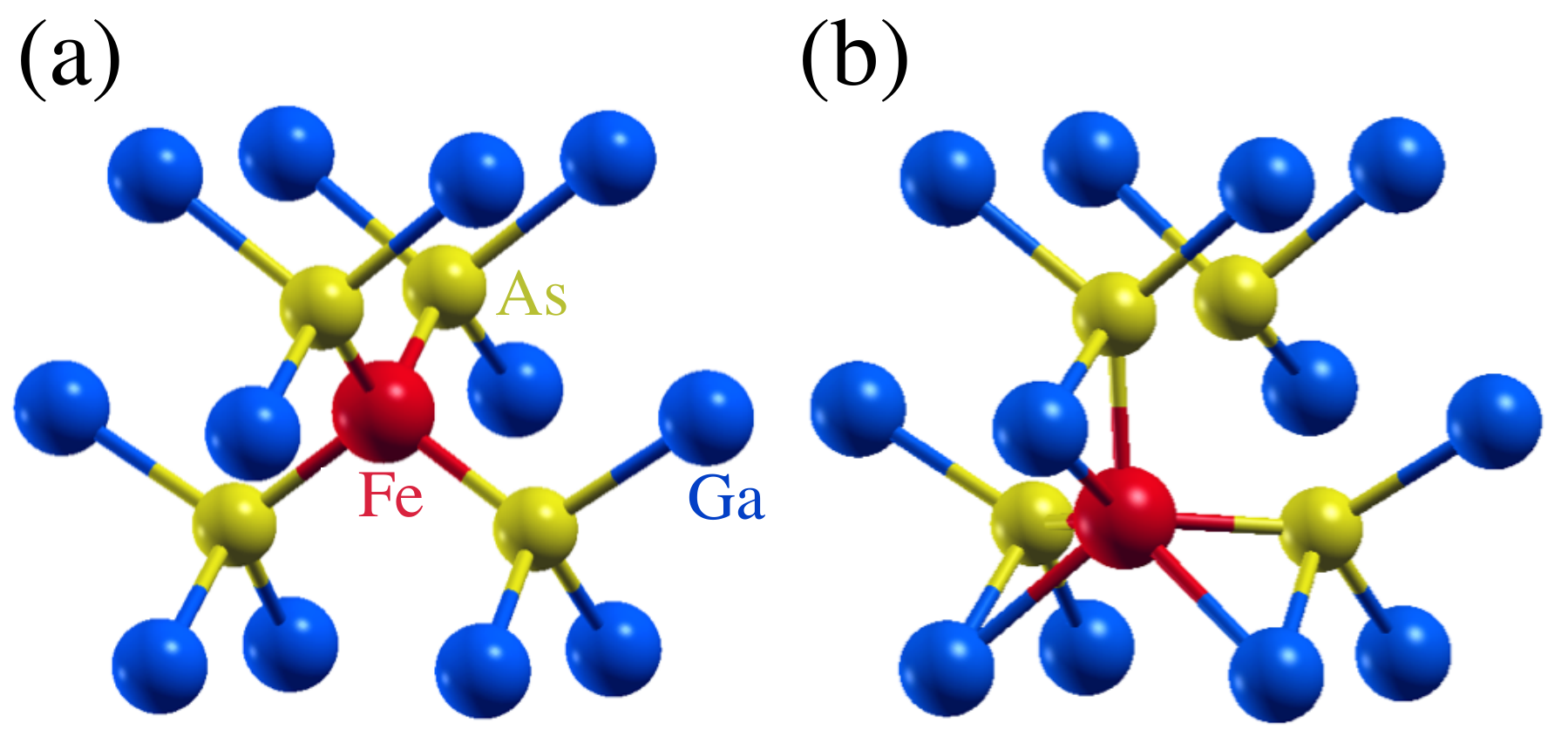}
\caption{\label{structs}(a) 4-fold symmetric substitutional lattice position of Fe in GaAs corresponding with type~I Fe.  (b) Interstitial lattice position of negatively charged Fe in GaAs corresponding with type~II Fe.}
\end{figure}

\begin{equation}
d^0 \rightarrow d^+ + e^-,
\end{equation}
 where $e$ is a free electron in the conduction band and the superscripts signify the charge state. The free electron then recombines with a different neutral impurity to form the $DX^-$ center,
\begin{equation}
d^0 + e^- \rightarrow DX^-.
\end{equation}
 where $DX$ is the broken-bond configuration. Thus, the complete reaction necessary for the formation of the $DX^-$ center is given by the equation
\begin{equation}
2d^0 \rightarrow d^+ + DX^-.
\end{equation}
 The formation energy for the case of a negatively charged structure must then take into account the addition of the free electron from the conduction band. This is included by amending the calculated total energies of the structure by the difference of the energy of the conduction band minimum relative to the Fermi energy.

In this work, we employ ground-state density functional theory (DFT) calculations using the linearized augmented plane wave (LAPW) method as implemented by the WIEN2k code\cite{wien2k}. The Perdew-Burke-Ernzerhof (PBE) generalized gradient approximation (GGA) exchange correlation functional was used. For each self-consistent calculation a 7x7x7 k-point grid was used and the wave functions were expanded in a plane wave basis set with a cutoff energy of 10 Ry. The evaluated structures consist of a 2x2x2 64-atom supercell of GaAs with a single dopant atom substituting in a Ga cation position. For the neutral and singly negatively charged cases, the dopant atom was displaced in increments of approximately 20 pm along the [111] axis away from the neighboring As atom to simulate the broken As bond, and the structure optimized at each step with the dopant position held fixed until the forces acting on each atom are less than 2 meV/\AA. For other charge states of Fe, the atom was displaced in increments of 40 pm along the [111] axis with the structure optimized at each step until the forces acting on each atom are less than 5 meV/\AA. For calculations performed with charged supercells, the final energies were amended using the Makov-Payne correction to first order~\cite{Makov1995}. The Makov-Payne correction eliminates artifacts which arise in \textit{ab initio} calculations with charged supercells due to the periodic boundary conditions implemented in the calculation. This is performed by an addition to the final energies of the form

\begin{equation}
E = E_0 - \frac{q^2\alpha}{2L},
\end{equation}

\noindent where $q$ is the additional charge, $\alpha$ is the lattice-dependent Madelung constant, and $L$ is the linear size of the supercell.

In order to calculate the effect of the shift from substitutional to interstitial lattice position on the Fe wavefunction, the \textit{ab initio} results have been used as input for a tight-binding calculation.  The tight-binding calculation, which is an extension of the approach of Ref.~\onlinecite{Tang2004}, uses an sp$^3$ model for the GaAs host with the addition of $d$ orbitals on the Fe dopant.  In order to include the effect of the shifted position of the Fe, we use the generalization of Harrison's $d^{-2}$ scaling law to calculate the new value of the hopping matrix element~\cite{Jancu1998} for the shifted position.  In the shifted case we amend the hopping between the Fe $d$ orbitals and the three nearest As $p$-orbitals from their non-shifted values based on the \textit{ab initio} calculated bond lengths.  Hopping from the Fe to the original fourth nearest neighbor As atom is taken to be zero, simulating a broken bond.

\begin{figure*}
\includegraphics[width=\textwidth]{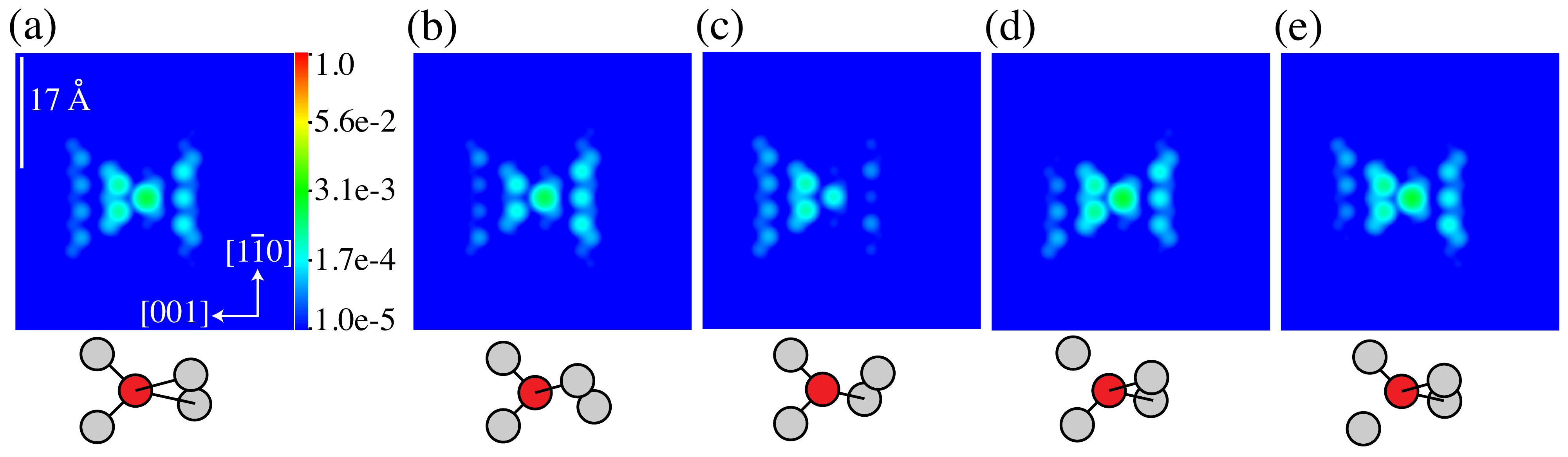}
\caption{\label{wavefunction} Wavefunctions of the t$_2$ symmetry state of Fe three monolayers above the position of the Fe for (a) the substitutional position and (b)-(e) interstitial position with different symmetries relative to the (110) plane.  The cartoons beneath the wavefunction plots depict the bond symmetry for each plot relative to the the (110) plane containing the Fe.}
\end{figure*}

\section{Results and Discussion}

The energy profiles in configuration space for the neutral and negative charge states of both Si and Fe can be seen in Fig.~\ref{energies}. Initially the calculations described in the previous section were performed for the case of Si in bulk GaAs for both charge states. The location of the second minimum of the formation energy of the negatively charged Si in the $DX^-$ center configuration was found to be at an approximate 1.37 \AA\ displacement from the 4-fold symmetric substitutional site along the [111] axis, while the interatomic distance from the broken-bond As increases from 2.40 \AA\ to 3.47 \AA\ resulting in a net interatomic displacement of 1.07 \AA\ from the neighboring As atom, in good agreement with previous calculations~\cite{Yi2011}. The $DX^-$ configuration of Si is one of 3-fold coordination symmetry. The formation energy of the broken-bond configuration, $E_f$, is calculated in the same manner as found in the work of Yi et al.~\cite{Yi2011} in which only energies between same charge states are compared. The formation energy is calculated from the equation

\begin{equation}
E_f = E(DX^-) - E(d^-),
\end{equation}

\noindent where $d^-$ is the energy of the negatively charged dopant atom at the 4-fold symmetric Ga substitutional position and $DX^-$ is the negatively charged dopant in the interstitial configuration of lowest total energy. Using this definition, the formation energy for Si in the $DX^-$ configuration was found to be 353 meV, suggesting a metastable configuration as expected for the case of in bulk GaAs. Previous calculations~\cite{Yi2011} have found that this formation energy switches from positive to negative as the Si atom is brought to the surface layer of [110] GaAs.

These calculations were repeated for the cases of neutral and negatively charged Fe in bulk GaAs. The resulting structures corresponding with the energy minima for type~I and type~II are pictured in Fig.~\ref{structs}.  For the case of negatively charged Fe, the location of the second minimum of total energy in configuration space corresponds to an approximate displacement of the Fe atom by 1.76 \AA\ from the 4-fold symmetric substitutional position along the [111] axis, while the interatomic distance between the Fe and broken-bond As increases from 2.32 \AA\ to 3.87 \AA\, resulting in a net increase of 1.55 \AA. The associated formation energy of this configuration was found to be 1.20 eV, implying that the interstitial configuration is only metastable in bulk GaAs, similar to the case of Si. The resemblance of the energy profiles for the Si and Fe cases in bulk GaAs is suggestive that the switching behavior of Si in bulk and near the surface may serve as an analogous model for the behavior of Fe at the surface. This would imply that the formation energy of the negatively charged Fe in the broken bond configuration may become positive as approaches the surface layer, a behavior which would be supported by the experimental observation of a change of topography near the ionized Fe dopant near the surface layer of [110] GaAs~\cite{Bocquel2013}.

\begin{figure}
\includegraphics[width=\columnwidth]{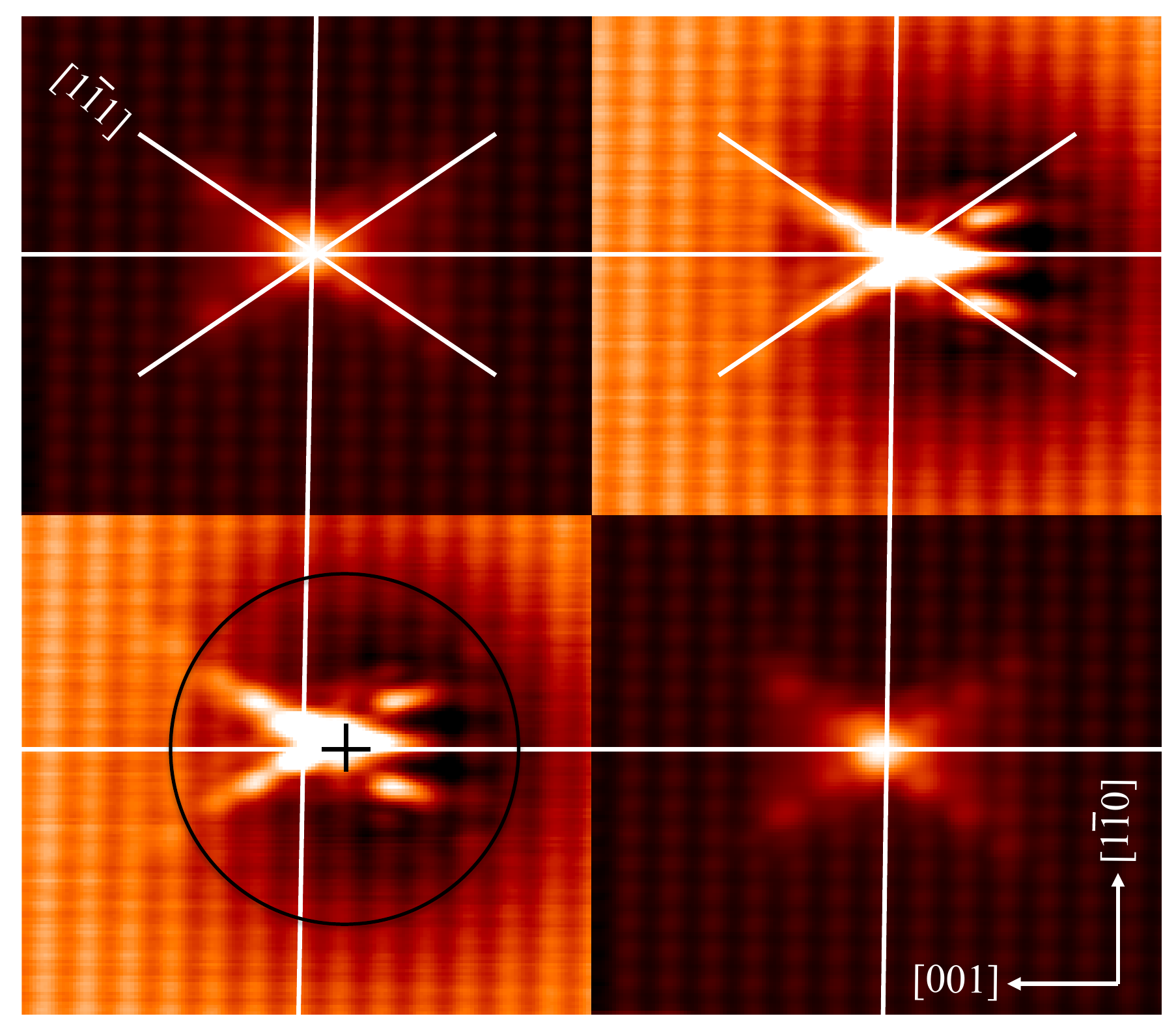}
\caption{\label{analysis} Topographic images of the same Fe atom in the type~I and type~II state. For both types the contrast is symmetric relative to the (110) plane containing the Fe. The strong contrast features appearing along the [111] directions cross each other in the same point for both type~I and type~II. This crossing point coincides with the position of a Ga atom in the surface of GaAs. The dark background related to the Coulomb field of atomically localized negative charge is shifted by about a lattice constant along the $[00\bar1]$ direction as indicated by the black circle and cross that are centered with respect to the dark background in the contrast.}
\end{figure}

One notable difference between the Fe and Si cases is the greater displacement of the Fe atom along the [111] axis by approximately 0.39 \AA, which places the atom into a configuration of 6-fold coordination symmetry as opposed to the 3-fold coordination symmetry seen with the Si $DX^-$ configuration. Another difference appears when comparing the neutral cases of Si and Fe. Whereas the neutral Si atom is completely unstable at all displacements away from the 4-fold symmetric substitutional configuration, the neutral Fe energy profile has a shallow second minimum, approximately 103 meV in depth, with a similar atomic displacement to the interstitial configuration for the negatively charged Fe. However, this second minimum for the neutral Fe has an associated formation energy of 1.79 eV, 590 meV greater than that of the negatively charged configuration.

The doubly negative and singly positive charge states of Fe in bulk GaAs both exhibit a second minimum, however they are found to be very shallow, similar to the case of neutral Fe, with depths of approximately 99 meV and 59 meV for the singly positive and doubly negative charge states respectively.  These energy minima correspond to displacements of the Fe along the [111] axis greater than 2 \AA. The formation energies are found to be 1.82 eV and 2.31 eV for the doubly negative and singly positive charge states respectively, with total energies greater than both the singly negative and neutral cases. These energy profiles suggest that there are no viable configurations of the Fe dopant along the [111] axis which exhibit an associated doubly negative or singly positive charge state, leaving the (Fe$^{2+}$)$^-$ acceptor state as the only metastable charge state for the interstitial Fe configuration.

The tight-binding wavefunction calculations further support  the origin of type~II Fe as due to a shifted lattice position of the Fe dopant.  In Fig~\ref{wavefunction}(a) the  bow~tie shape of the type~I (substitutional) Fe t$_2$ symmetry state is clear.  In Fig~\ref{wavefunction}(b)-(c) the bond between the Fe and a nearest neighbor As below/above the (110) plane containing the Fe has been broken and the Fe shifted towards its nearest neighbors in the [001] direction.  In Fig~\ref{wavefunction}(d)-(e) the bond between the Fe and one of its nearest neighbor As in the (110) plane has been broken and the Fe shifted closer towards its other nearest neighbors.  In the tight-binding calculation the largest change to the bonding configuration between type~I and type~II Fe is the broken bond between the Fe and one of its nearest neighbor As, which does not destroy the overall bow~tie shape of the Fe wavefunction.

Close inspection of the topographic contrast for the same Fe atom in either the type~I or type~II state shows that the bright and dark contrast features are symmetric relative to the (110) plane containing the Fe. The topographic images furthermore show that the strong contrast appearing along the [110] directions, especially on the lefthand side of the images, cross in a position that coincides with a Ga atom at the surface. This can be understood if the Fe atom is located in an odd layer below the semiconductor/vacuum interface. We estimate that the Fe atom is in the 3rd layer counting from the semiconductor/vacuum interface. The symmetry center of the dark contrast that we relate to negative charge that is localized at the atomic scale is shifted by about one lattice constant along the $[00\bar1]$ direction. We suggest that this negative charge is the due to the electrons that are   captured in the broken bonds and this charge is shifted away from the Fe atom. Taking into account the symmetry of the bright contrast features that are related to wavefunction that has been calculated for various geometries  we can conclude that the broken bond is aligned along the [001] direction. This corresponds with the geometries for the wavefunction calculations shown in Fig~\ref{wavefunction}(b) or (c). It is quite likely that the presence of the surface favors a broken bond along this symmetry direction~\cite{Garleff2011,Smakman2014}.

\section{Conclusions}

The results of our \textit{ab initio} calculations show that Fe has a second metastable configuration in bulk GaAs with a formation energy of 1.2 eV with a configuration similar\cite{Chadi1989} to the $DX^-$ center observed in the bistable behavior of donors in Al$_x$Ga$_{1-x}$As alloys for $x\geq0.22$. This second configuration of Fe differs from the $DX^-$ center by its larger displacement from the Ga substitutional position to an interstitial position of six-fold coordination symmetry. This second metastable configuration is accompanied by an associated change of charge state from the isoelectronic (Fe$^{3+}$)$^0$ state to the (Fe$^{2+}$)$^-$ acceptor state. Our results for the case of Si in bulk GaAs also predict metastable behavior with an associated change in charge state, which is in agreement with the results of previous first-principles investigations~\cite{Yi2011} of Si in bulk GaAs.  The tight-binding wavefunction calculations suggest that the change in bond configuration required to shift from type~I to type~II Fe does not eliminate the bow~tie shape of the wavefunction seen experimentally for both types.

Previous work investigating the bistable behavior of Si at the [110] surface of GaAs~\cite{Garleff2011} has demonstrated that surface effects can be very important in the properties of individual dopants, causing very different behavior from that seen in bulk. In the case of Si at the surface of GaAs, it has been shown in first-principles calculations~\cite{Yi2011} that the formation energy of the $DX^-$ center configuration of Si decreases to become more energetically favorable than the substitutional configuration if the atom resides in the top layer. This work provides evidence that a similar behavior occurs for case of Fe when the dopant is close to the surface layer and the negative charge state is electronically induced by STM. The second metastable configuration coordinate of Fe in bulk GaAs along with these experimental findings would imply that the bistability behavior of Fe is analogous to the case of Si donors in GaAs in which surface effects play a large role in the formation of the defect. The bistability of Fe at the surface of GaAs implies the possibility to electronically manipulate both the charge state and lattice position of a TM dopant using STM.

\acknowledgments

This work was supported by an AFOSR MURI.

J. M. M. and V. R. K. contributed equally to this work.

\end{document}